\input harvmac
\input epsf
%\draftmode

\font\ticp=cmcsc10
 
\def\Title#1#2{\rightline{#1}\ifx\answ\bigans\nopagenumbers\pageno0\vskip1in
\else\pageno1\vskip.8in\fi \centerline{\titlefont #2}\vskip .5in}

\font\ticp=cmcsc10

\def\ie{{\it i.e.,}\ }
\def\eg{{\it e.g.,}\ }

\def\l{\ell}

\def\lp{\l_{\rm fun}}

\lref\rEHM{R.~Emparan, G.T.~Horowitz and R.C.~Myers, 
J. High Energy Phys. {\bf 01} (2000) 007, 
%{\it
%``Exact Description of Black Holes on Branes,"} 
hep-th/9911043.}
\lref\rSUS{L.~Susskind, private communication}
\lref\rADD{N.~Arkani-Hamed, S.~Dimopoulos, and G.~Dvali, Phys. Lett.
{\bf B429} (1998) 263, hep-ph/9803315;
Phys.\ Rev.\  {\bf D59} (1999) 086004, hep-ph/9807344;
I.~Antoniadis, N.~Arkani-Hamed, S.~Dimopoulos and G.~Dvali,
Phys.\ Lett.\  {\bf B436} (1998) 257, hep-ph/9804398.}
\lref\rBF{T.~Banks and W.~Fischler, {\it ``A Model for High Energy Scattering
in Quantum Gravity"}, hep-th/9906038.}
\lref\rADM{P.~Argyres, S.~Dimopoulos, and J.~March-Russell, Phys. Lett. 
{\bf B441} (1998) 96, hep-th/9808138.}
\lref\rRS{L.~Randall and R.~Sundrum,
Phys.\ Rev.\ Lett.\  {\bf 83} (1999) 4690, hep-th/9906064;
3370, hep-ph/9905221.}
\lref\rGL{R.~Gregory and R.~Laflamme,
Phys.\ Rev.\ Lett.\ {\bf 70} (1993) 2837,
hep-th/9301052; Nucl.\ Phys.\  {\bf B428} (1994) 399, hep-th/9404071.}
\lref\rGKR{S.~Giddings, E.~Katz, and L.~Randall, {\it ``Linearized Gravity
in Brane Backgrounds,"} hep-th/0002091.}
\lref\rCHR{A.~Chamblin, S.W.~Hawking and H.S.~Reall,
Phys.\ Rev.\  {\bf D61} (2000) 065007, hep-th/9909205.}
\lref\rEHMII{R.~Emparan, G.~T.~Horowitz and R.~C.~Myers,
%{\it ``Exact Description of Black Holes on Branes II: 
%Comparison with BTZ Black Holes and Black Strings,''}
J. High Energy Phys. {\bf 01} (2000) 021, 
hep-th/9912135.}
\lref\rWITT{B.S.~DeWitt, Phys. Rep. {\bf C19} (1975) 297.}
\lref\rPAGE{D.~Page, Phys. Rev. {\bf D13} (1976) 198; Phys. Rev. D{\bf
14} (1976) 3260.}
\lref\rSANCHEZ{N.~S\'anchez, Phys. Rev. {\bf D18} (1978) 1030;
pages 461-479 (hep-th/9711068) in `String Theory in Curved Spacetimes,' ed.,
N.~S\'anchez (World Scientific, 1996).}

\baselineskip 16pt
\Title{\vbox{\baselineskip12pt
\line{\hfil hep-th/0003118}}}
{\vbox{
{\centerline{Black Holes Radiate Mainly on the Brane}}
}}
\centerline{\ticp Roberto Emparan$^a$, Gary T. Horowitz$^b$,
Robert C. Myers$^c$}
\vskip 2ex
\centerline{\it $^a$ Departamento de F{\'\i}sica Te\'orica,
Universidad del Pa{\'\i}s Vasco, Apdo.\ 644, E-48080 Bilbao, Spain}
\vskip 1ex
\centerline{\it $^b$ Physics Department, University of
California, Santa Barbara, CA 93106 USA}
\vskip 1ex
\centerline{\it $^c$ Department of Physics, McGill University, Montr\'eal,
QC, H3A 2T8, Canada}
\vskip 2ex
\centerline{$^a$wtpemgar@lg.ehu.es, $^b$gary@cosmic.physics.ucsb.edu,
$^c$rcm@hep.physics.mcgill.ca}

\vskip 2cm
\centerline{\bf Abstract}
\bigskip
We examine the evaporation of a small black hole on a brane in a world with
large extra dimensions. Since the masses of many Kaluza-Klein modes are much
smaller than the Hawking temperature of the black hole, it has been claimed
that most of the energy is radiated into these modes. We show that this is
incorrect. Most of the energy goes into the modes on the brane. This raises
the possibility of  observing Hawking radiation in future high energy colliders
if there are large extra dimensions.

\Date{March, 2000}

\newsec{Introduction}

It has been proposed that space may have extra compact dimensions as
large as a millimeter \rADD. If all the standard model fields live on a
three-brane and only gravity (and perhaps some other unobserved fields)
propagate in the bulk, such large extra dimensions are consistent with
all current observations. We will consider the evaporation of black
holes in this scenario. Although our results hold for any number of
large extra dimensions, for definiteness we focus mainly on the case of
two extra dimensions of size $L$. Since the effective four-dimensional
Newton's constant $G_4$ is related to $G_6$ by $G_4 = G_6/L^2$, if the
fundamental scale of gravity in the bulk is of order a TeV, $G_4$ has
the observed value provided $L\sim 1$~mm. For weak fields, the bulk
metric can be decomposed into the four-dimensional graviton and an
infinite tower of Kaluza-Klein modes, which act like four-dimensional
spin-two fields with masses starting at $1/L\sim 10^{-4}$~eV.

One of the most striking consequences of a low fundamental Planck scale,
is the possibility of forming semiclassical black holes at rather low
energies, say of order 100~TeV.
Suppose one collapses matter (or collides particles) on the brane to
form a  black hole of size $\lp \ll r_0 \ll L$
(where $\lp=G_6^{1/4}$ is the fundamental, \ie six-dimensional,
Planck length). This black hole has a temperature
$T\sim 1/r_0$ which is much larger than the mass of the light Kaluza-Klein
modes. Since gravity couples to everything, and there are so many Kaluza-Klein
modes with mass less than the Hawking temperature, it has been claimed
\refs{\rADM, \rBF}
that the Hawking radiation will dominated by these Kaluza-Klein modes, with only a tiny
fraction of the energy 
going into standard model particles. In other words,
most of the energy would be radiated off of the brane into the bulk.
If this were the
case, the Hawking radiation from these small black holes
would be essentially unobservable. 

We claim that this argument is incorrect, and most of the Hawking radiation
goes into the standard model fields on the brane! The easiest way to see this
is to consider the calculation from  the six-dimensional perspective\foot{This
argument was given in a slightly different context in \rEHM. Similar
observations were also made independently by Susskind \rSUS.}. For a single
massless six-dimensional
field, the rate at which is energy radiated is of order
\eqn\hawsix{ {dE\over dt} \sim A_6 T^6 \sim {r_0^4 \over r_0^6} 
\sim {1\over r_0^2} }
where $A_6$ denotes the 
area of the six-dimensional black hole.
For a single massless four-dimensional field on the brane, the rate of energy
loss is of order
\eqn\hawfour{ {dE\over dt} \sim A_4 T^4 \sim {r_0^2 \over r_0^4} 
\sim {1\over r_0^2} }
and hence is the same. That is, with a single relevant scale $r_0$
determining the Hawking radiation, bulk and brane fields must 
both have $dE/dt\sim r_0^{-2}$. 
Hence the Hawking evaporation must 
emit comparable amounts of energy into the bulk and brane. However,
with the typical assumption that there are
many more fields on the brane than in the bulk,
one would conclude that most of the energy
goes into the observable four-dimensional
fields. While the detection of this Hawking radiation
would likely not be the first experimental signature of large extra
dimensions, such measurements would provide a dramatic new window
on black hole microphysics. 

We will examine this argument in more detail below (and confirm its
validity), but first we must ask what was wrong with the original
arguments suggesting that the Hawking radiation goes mostly into
Kaluza-Klein modes. In one form \rADM, one views the emission of
Hawking radiation as a six-dimensional process. In this case, since
brane fields seem to have a tiny phase space compared to bulk fields, it
would appear that the emission of the latter should dominate the Hawking
evaporation. However, it is incorrect to think of brane fields as bulk
fields confined to a limited phase space. The brane fields are
intrinsically four-dimensional, and their emission is governed by the
four-dimensional relation \hawfour, and {\it not} the six-dimensional
formula \hawsix\ with a restricted area.

Dominance of the Kaluza-Klein modes might also be argued from a four-dimensional
point of view \rBF. In this case, it may appear that the Kaluza-Klein modes
must dominate the evaporation since there are a large number (of order
$(L/r_0)^2$) light
modes with masses below the scale of the Hawking temperature.
However, here it is incorrect to think of the individual
Kaluza-Klein modes of the bulk graviton as massive spin two fields on the brane
with standard (minimal) gravitational
couplings. Rather, since the Kaluza-Klein modes are excitations in the full
transverse space, their overlap with the small (six-dimensional) black
holes is suppressed by the geometric factor $(r_0/L)^2$ relative to the
brane fields.
Hence this geometric suppression precisely compensates for the enormous
number of modes, and the total contribution of all Kaluza-Klein modes is only the
same order as that from a single brane field. Since eq.~\hawsix\
automatically incorporates the emission of all Kaluza-Klein modes, clearly 
this four-dimensional approach is a complicated reorganization of a simple
six-dimensional situation.

\newsec{Detailed calculations}

We now want to look in more detail at the rate of energy loss by
a black hole to modes on the brane and in the bulk.
We will consider a general dimension $d$ for the bulk spacetime, and
assume that we live on a (3+1)-dimensional brane. The extra dimensions will 
have size $L$.
Since we are assuming the size of the black hole $r_0$ is much less than $L$,
the geometry near the black hole is simply that of a $d$-dimensional 
Schwarzschild solution
\eqn\metric{
ds^2=-f(r)\,dt^2+{dr^2\over f(r)}+r^2d\Omega_{d-2}^2}
with 
\eqn\deff{f(r)=1-\left({r_0\over r}\right)^{d-3}.}
The event horizon is thus at $r=r_0$,
and the area of the event horizon is $A_d=r_0^{d-2}\Omega_{d-2}$
where $\Omega_n$ denotes the volume of a unit $n$-sphere.

If a black hole is formed from matter on the brane, symmetry requires
that the brane pass through the equator of the black hole.
We further assume that the three-brane is essentially a test brane with
negligible self gravity of its own\foot{We also assume that the brane
has negligible thickness. This is reasonable since the actual thickness of the
brane is likely to be of order the fundamental
scale $\lp$, and a black hole will behave semi-classically only if $r_0
\gg \lp$.}.
Then the induced metric on the brane will be
\eqn\inmetric{
ds^2=-f(r)\,dt^2+{dr^2\over f(r)}+r^2d\Omega_2^2}
with $f(r)$ still given by \deff. On the brane then, the event horizon is again
at $r=r_0$, and the area of the event horizon is $A_4=4\pi r_0^2$.
This induced metric on the brane 
is certainly not the four-dimensional Schwarzschild
geometry. 
Since the Ricci tensor of this four-dimensional metric \inmetric\ 
is nonzero near the horizon, one can think
of it as a black hole with matter fields (\ie Kaluza-Klein modes) around it. 
However, the calculation of Hawking
evaporation relies mainly on properties of the horizon, such
as its surface gravity. Changing the geometry outside will change the
effective potential that waves have to propagate through. 
This will  modify the
grey body factors, but since the potential is qualitatively the same,
 the total
energy radiated is changed only by factors of order unity.
Since the Hawking temperature is constant over the horizon, it is the same
for both 
the black hole in the bulk and on the brane,
and is given by
\eqn\temp{
T={d-3\over 4\pi r_0}\ .}

The metric \inmetric\ (with $f$ given by \deff)
has no $1/r$ term and hence seems to give zero mass in four dimensions.
However,
this metric only describes the geometry near the black hole. For $r\gg L$
the geometry will be approximated by \inmetric\ with 
\eqn\newf{f(r)\simeq 1-{2G_4 M\over r}}
where $M$ is the mass of the $d$-dimensional black hole
\eqn\mass{M= {(d-2)r_0^{d-3} \Omega_{d-2}\over 16\pi G_d}}
In other words, the mass measured on the brane is the same as the mass
in the bulk.
This can be seen as follows. Consider the higher dimensional spacetime and
unwrap the compact dimensions. The result is a cubic array of black holes, each
of mass $M$ and separated by a distance $L$. From a large
distance,\foot{Here, we ignore the gravitational interaction energy
of the black holes in the array, which is justified for $r_0\ll L$.} this
looks like a ``surface density" $\rho =M/L^{d-4}$. The asymptotic metric will
thus contain the term $f(r) = 1- (2G_d \rho/r)$. However, since
$G_d = G_4 L^{d-4}$,
this is equivalent to \newf. Although this $1/r$ term is the dominant
correction to the flat metric for $r\gg L$, it is already quite small
for $r\sim L$ and will not cause a significant modification to our
estimates of the energy radiated. 

We now show that
the emission rate of Kaluza-Klein modes, regarded
as four-dimensional fields, is actually suppressed relative to modes
that propagate only along the brane. In order to see this, let us
consider the calculation of the emission rate of a  massless
bulk field in the
following way: since we have to sum over all the modes of the field that
are emitted by the black hole, let
us decompose these according to the  momentum ${\bf k}$ which they carry into
the $d-4$ transverse dimensions. On the brane, this Kaluza-Klein
momentum is identified with the
{\it four-dimensional} mass of these modes, which we denote $m=|{\bf k}|$.
If we then sum over all other quantum
numbers, we will find the emission rate corresponding to a Kaluza-Klein mode with
momentum ${\bf k}$. Proceeding in this way we get, for
the emission rate per unit frequency interval, of modes with momenta in
the interval (${\bf k}, {\bf k}+ d{\bf k}$),
\eqn\ratekk{
{dE\over d\omega dt}(\omega,{\bf k})\simeq
(\omega^2-m^2) {\omega A_d
\over e^{\beta\omega}-1}\,d^{d-4}\!k\ .
}
Here, $A_d$ is the area of the black hole
in the $d$-dimensional bulk\foot{The
only difference for a {\it fermionic} mode would, of course,
be to change the sign of the `one'
in the denominator in the formula above.}. We are
neglecting purely numerical factors since we will find below that they
do not play
any significant role. As a check, when
this expression is integrated over all Kaluza-Klein modes, one recovers
the emission rate of a massless bosonic field into the $d$-dimensional
bulk:
\eqn\bulkrate{
{dE\over d\omega dt}(\omega)=\int_{|{\bf k}|=0}^{|{\bf k}|=\omega}
{dE\over d\omega
dt}(\omega,{\bf k}) \ d^{d-4}\!k\simeq {
\omega^{d-1} A_d\over
e^{\beta\omega}-1}\,.
}
 %Now 
Consider a light Kaluza-Klein mode, with a mass
much smaller than the black hole temperature, 
$m \ll 1/r_0$. 
We set $d^{d-4}\!k\sim (1/L)^{d-4}$ for an individual mode, and  
$A_d\sim r_0^{d-4} A_b$, with $A_b$ the sectional area on the brane.
Then, 
\eqn\lightkk{
{dE\over d\omega dt}(\omega,m)\simeq \left( r_0\over L\right)^{d-4}
(\omega^2-m^2) {\omega A_b\over
e^{\beta\omega}-1}\,.
}
which is identical to the emission rate of a massive field in four dimensions,
except for a suppression factor of $(r_0/L)^{d-4}$. (Note that this formula
applies equally well for $m=0$).
So we see that the Hawking radiation into each 
Kaluza-Klein mode (among these, the massless graviton) is much smaller
than the radiation into any other minimally coupled field that
propagates only in four dimensions. In particular, compared with a
purely four-dimensional gravity theory, Hawking radiation in gravitons
on the brane is suppressed by a factor of $(r_0/L)^{d-4}$. 
Still the total radiation \bulkrate\ into a bulk field is comparable to that
into a field on the brane, because there
are of order $(L/r_0)^{d-4}$ light modes with $m<T\sim1/r_0$.
As we mentioned earlier, this
suppression factor can be understood as arising from the small geometric
overlap between a bulk mode and a small black hole which has only a
limited extent in the transverse dimensions. Of course, 
since there is no analogous effect for all the nongravitational fields on
the brane, this supports our conclusion that most of the energy is radiated
on the brane. 

Since the number of relevant fields on the
brane may be only a factor of ten or so larger than the number of bulk
fields, one might worry that the claim that the Hawking radiation is
dominated by brane fields could still be thwarted by large numerical
factors coming from the higher dimensional calculation. To check this,
we consider two improvements over the rough estimate of the radiation
rates given in \hawsix\ and \hawfour.
The first is to include the
dimension dependent Stefan-Boltzman constant $\sigma_n$. In $n$ dimensions,
the energy radiated by a black body of temperature $T$ and surface area
$A_n$ is
\eqn\stefan{
{dE_n\over dt}= \sigma_n\, A_n\, T^n}
Repeating the standard calculations found in any statistical mechanics
text, in higher dimensions, one finds that the $n$-dimensional
Stefan-Boltzman constant is
\eqn\constant{\eqalign{
\sigma_n=&{\Omega_{n-3}\over(2\pi)^{n-1}(n-2)}\int^\infty_0
{z^{n-1}\,dz\over e^z-1}\cr
=&{\Omega_{n-3}\over(2\pi)^{n-1}(n-2)}\Gamma(n)\,\zeta(n)\cr}}
with $\zeta(n)$ denoting the Riemann zeta function. These factors do not 
change much with the dimension, in the cases of interest. For
example,
\eqn\svalues{
\sigma_4={\pi^2\over 120}\simeq .08,\quad
\sigma_6={\pi^3\over 504}\simeq .06,\quad
\sigma_{10}={\pi^5\over 3168}\simeq .097}
Although formally these quantities have been calculated for infinite
(uncompactified) spacetimes, eq.~\stefan\ provides a good approximation 
when $T\gg 1/L$. 
The fact that $\sigma_n$ changes very little with dimension, confirms that even
though higher dimensional spacetimes have infinitely many more modes
(corresponding to excitations in the extra dimensions), the rate
at which energy is
radiated by a black body with radius $r_0$ and temperature $T\sim 1/r_0$
is roughly independent of the dimension.

Substituting eq.~\constant\ in eq.~\stefan, we find for the black hole that
\eqn\stefanc{
{dE_n\over dt}=\sigma_n \Omega_{n-2} 
\left({d-3\over4\pi}\right)^n {1\over r_0^2}
={\Omega_{n-3}\over(2\pi)^{n-1}(n-2)}\Gamma(n)\,\zeta(n)
\,\Omega_{n-2} \left({d-3\over4\pi}\right)^n {1\over r_0^2}}
where we have used the horizon area for $A_n$.
Hence for modes in a three-brane, we find
\eqn\four{
{dE_4\over dt}={(d-3)^4\over 7680\pi} {1\over r_0^2}\ .}
For the 
case of a six-dimensional world, $n=6$, with two extra large compact
dimensions 
\eqn\six{
{dE_6\over dt}={(d-3)^6\over 4^6\cdot 189\pi} {1\over r_0^2}\ .}
Now if we substitute in $d$=6 and take the ratio, we find
\eqn\ratio{
{dE_4/dt\over dE_6/dt}={56\over5}= 11.2\ .}
Hence by these calculations, the emission of a bulk mode is actually
suppressed relative to a mode confined to the brane.
If we consider $n$=$d$=10, the ratio becomes
\eqn\ratiob{
{dE_4/dt\over dE_{10}/dt}\simeq 12.1\ .}

However, there is a second improvement which we can easily incorporate
into our calculations.
This concerns the area that appears in \stefan.
We have been using the horizon area as the  area of the black body
emitter in eq.~\stefan, but at least in the geometric optics approximation,
a black hole acts as a perfect absorber of a slightly larger radius.
Recall that in four dimensions, there is a critical radius $r_c=(3\sqrt{3}/2)
r_0\simeq 2.6\, r_0$ for null geodesics. If a photon travels inside this radius,
it is captured
by the black hole. Detailed calculations have shown \rSANCHEZ\ that 
the total energy radiated is better approximated
by assuming the area is given by $r_c$ rather than $r_0$. Note,
however, that this DeWitt approximation \rWITT\ is not obviously
justified since the typical wavelengths are of order the
size of the black hole.

Although detailed calculations are not yet available in higher dimensions,
we expect a similar improvement exists in this case as well. 
For a general dimension, $r_c$ becomes
\eqn\crit{
r_c=\left({d-1\over 2}\right)^{1/(d-3)}\sqrt{d-1\over d-3}r_0\, .}
The ratio decreases slightly with the dimension:
at $d=6$, $r_c\simeq 1.75\, r_0$;
at $d=10$, $r_c\simeq  1.41\, r_0$.
Note that this critical radius will be the same for brane and bulk
modes since the problem of calculating null geodesics involves only
motion in a plane of the full geometry \metric.
The correction due to this effect enters the emission rate \stefanc\
through the area factor. Since the bulk modes include a higher power of
the radius, increasing the radius increases the relative decay rates
for the bulk modes by a
factor $(r_c/r_0)^{n-2}$. With this correction, we find
\eqn\ratiocorr{
{dE_4/dt\over dE_6/dt} %\simeq 11.2\times(1.75)^{-2}
\simeq 3.66\ ,\qquad
{\rm and}\qquad
{dE_4/dt\over dE_{10}/dt} %\simeq 12.1\times(1.41)^{-6}
\simeq 1.54\ ,}
and so the ratios become closer to one.

Thus there are no unexpected large factors to ruin the
naive estimate that a Hawking evaporation emits as much energy
into a typical brane field as into a typical bulk field. A definitive
comparison of the bulk and brane radiation rates would require a more detailed
analysis. In particular, one expects a suppression for higher spin fields
due to angular momentum barriers \rPAGE.  For example, in a pure
four-dimensional calculation, the radiation rate for the graviton is
approximately 10 times smaller than that for a massless spin-one-half
field \rPAGE. Of course, such detailed calculations would require a
specific brane-world model to determine the exact black hole geometry
and the precise multiplicity of bulk and brane fields.

\newsec{Discussion}

So far we have considered small black holes with $r_0 < L$. 
Will larger black holes also radiate mainly on the brane? If $r_0 >L$, the
solution is simply a product of four-dimensional Schwarzschild and a torus.
Hence the horizon area is $A_d = 4\pi r_0^2\, L^{d-4}$, and the geometric
suppression factor in eq.~\lightkk\ is replaced by one. However,
the Hawking temperature is now lower than the mass of all Kaluza-Klein
modes, so their contribution to the Hawking radiation is clearly suppressed.
Approximating the radiation rate with eq.~\hawsix, we have
\eqn\newrate{
{dE\over dt}\sim A_d T^d \sim r_0^2 T^4 (LT)^{d-4}\sim (L/r_0)^{d-4} r_0^2 T^4 
\ .}
So the total contribution of the Kaluza-Klein modes is suppressed by the factor
$(L/r_0)^{d-4}$ relative to that a single brane field.
Actually, since $T<1/L$,
this six-dimensional formula only accurately captures the
contributions of modes with relatively large Kaluza-Klein momentum.
The dominant contribution will actually come from the massless mode
which in this regime radiates identically to a brane field.
So for large black holes, a bulk field still carries essentially the same
energy as a  field on the brane, and the latter again dominate
the Hawking radiation due to the relatively high multiplicity of light
brane fields.

If a black hole initially has $r_0 >L$, then Hawking radiation will
cause the Schwarzschild radius to decrease. When $r_0\sim L$, the
four-dimensional black hole $\times$ $(S^1)^{d-4}$ solution becomes
unstable \rGL, and is believed to break up into $d$-dimensional black
holes.\foot{Note that at the transition with $r_0 \sim L$, the 
black hole mass is $M \sim L^{d-3}/G_d = L/G_4$. Although
this is much larger than the four dimensional Planck mass, it is much smaller
than a typical stellar mass (\eg\ for $d=6$ and $L=1$ mm, $M=10^{27}$ gms,
which is about the mass of the Earth).}
These black holes attract each other
and coalesce, forming a single higher dimensional black hole. Could this
final black hole lie in the bulk and not on the brane? This is highly
unlikely since a black hole will {\it not} slide off a brane! Rather
it feels a
restoring force due to the brane tension. To see this, we must consider
the condition for a black hole on a brane to be static.
A black hole will grow
whenever $T_{\mu\nu} \l^\mu \l^\nu >0$ where $\l^\mu$ is a null
geodesic generator of the event horizon. This is just the statement that
energy is crossing the horizon. The stress energy tensor of a brane is 
proportional to its induced metric. In order for the black hole to be static
(and not swallow up the brane) $\l^\mu$ must lie entirely in the brane so
$T_{\mu\nu} \l^\mu \l^\nu \propto \l_\mu \l^\mu =0$. This will be the case
if the radial direction orthogonal to the black hole is tangent to the
brane. In other words, the brane must intersect the black hole orthogonally.
So if one pulls on a black hole on a brane, the brane bends to stay
orthogonal and pulls back on the black hole. Thus,
a black hole on the brane will attract a black hole in the bulk,
forming a larger black hole on the brane.

Although we have found that most of the radiation goes into purely 
four-dimensional fields, the evaporation of a small
black hole will not proceed as in a purely four-dimensional theory.
The black hole is 
$d$-dimensional, and its mass $M$ is related to the radius as in
\mass. In particular, this means that the lifetime of the black hole
will not be like that of a four-dimensional black hole, $\tau_4\sim G_4^2
M^3$, but rather, $\tau_d\sim G_d^{2\over d-3} M^{d-1\over d-3}$ \rADM.
Note that $\tau_d\sim (L/r_0)^{2(d-4)} \tau_4$ and so the lifetime
is longer (possibly enormously longer) than would have been
expected from four-dimensional Einstein
gravity. The essential feature is that when $G_dM<L^{d-3}$
(\ie $r_0<L$), for a fixed mass, the Schwarzschild radius is larger
than it would be for a four-dimensional black hole. This means that 
the temperature is lower, the horizon area is larger, and the evaporation rate
is slower.
The fact that the horizon area is larger is the  
feature which results in the higher dimensional black hole being
entropically favored \rGL. 
In the scenario with
$d=6$ and $L\simeq 1$~mm, the lifetime of a black hole formed at
$M\simeq 100$~TeV (so $r_0\sim 10^{-15}$~mm) would be $\tau_6
\sim 10^{-25}$~s.\foot{For a black hole with mass smaller than $10^{19}$~GeV,
it is not meaningful to compare its lifetime with a semiclassical
four-dimensional estimate.}

Finally, although we have focused our discussion 
on the large extra dimension scenario, black holes still radiate mainly
on the brane in the Randall-Sundrum scenario \rRS\ with an
infinite extra dimension. As discussed
in \refs{\rEHM,\rCHR,\rGKR,\rEHMII}, large black holes on the brane
(with Schwarzschild radius $r_0$
larger than the
scale $R$ of the bulk cosmological constant) appear as flattened pancakes
and have a five-dimensional area of order $A \sim r_0^2 R$. The temperature
is constant over the horizon and of order $T\sim 1/r_0$. So the energy
radiated in five-dimensional bulk modes is $dE/dt \sim A_5T^5 \sim R/r_0^3$
which is much smaller than the energy radiated in four-dimensional modes
on the brane: $dE/dt \sim A_4 T^4 \sim 1/r_0^2$
\rEHM. Black holes which are smaller than the AdS curvature scale will be
approximately spherical and behave as we have discussed above.

Given that small black holes radiate mainly on the brane (and that such
black holes will not slip off the brane), the brane-world scenario
has the potential to make interesting observable predictions about
small black holes appearing either in collider experiments or in the
early universe. It will be interesting to investigate 
their detailed phenomenology.

%\vskip 2cm

\bigbreak\bigskip\bigskip\centerline{{\bf Acknowledgements}}\nobreak

\vskip .5cm

The work of RE is partially supported by UPV grant 063.310-EB187/98 and
CICYT AEN99-0315. GTH was supported in part by NSF Grant PHY95-07065.
RCM was supported in part by NSERC of Canada and Fonds du Qu\'ebec. This
paper has report numbers EHU-FT/0003 and McGill/00-10. 

\listrefs

\end